\documentstyle[12pt,aaspp4,flushrt]{article}
\def\etal{{et~al.}}

\def\cf{{cf.}}
\def\HST{{\it Hubble Space Telescope}}

\def\Hii{{H~{\sc ii}}}

\def\arcsec{^{\prime\prime}}

\def\kpc{\mbox{~kpc}}
\def\Gyr{\mbox{~Gyr}}
\begin{document}

\title{Dynamics of ``Small Galaxies'' in the Hubble Deep Field}

\author{Wesley N. Colley, \footnote{Supported by the
Fannie and John Hertz Foundation, Livermore, CA 94551-5032}
Oleg Y. Gnedin, Jeremiah P. Ostriker, James E. Rhoads
\footnote{Present Address:  Kitt Peak National Observatory}}
\affil{Princeton University Observatory, 
Princeton, NJ 08544}
\begin{center}
Email: wes,ognedin,jpo,rhoads@astro.princeton.edu
\end{center}

\begin{abstract}

We have previously found in the Hubble Deep Field a significant angular
correlation of faint, high color-redshift objects on scales below one
arcsecond, or several kiloparsecs in metric size.  A correlation at this scale
is most likely due to physical associations.  We examine the correlation and
nearest neighbor statistics to conclude that 38\% of these objects in the HDF
have a companion within one arcsecond (or about $6\kpc$), three times the
number expected in a random distribution with the same number of objects; the
total excess approaches 1.5 objects by separations of 10 arcseconds.  We next
examine three possible dynamical scenarios for these object multiplets: 1) the
objects are star-forming regions within normal galaxies, whose disks have been
relatively dimmed by $K$-correction and surface brightness dimming; 2) they are
fragments merging into large galaxies; 3) they are satellites accreting onto
parent normal $L_*$ galaxies.  We find that hypothesis 1 is most tenable.
First, large galaxies in the process of a merger formation would have
accumulated too much mass in their centers ($5 \times 10^{12}M_\odot$ inside
$2~\mbox{kpc}$) to correspond to any abundant category of present day objects.
Second, accretion by dynamical friction occurs with a predictable slope in
density vs. radius that is not seen among the faint HDF objects.  Since the
dynamical friction time is roughly (1~Gyr), a steady-state should have been
reached by redshift $z \la 5$.  In the context of these two dynamical
scenarios, we consider the possible effects of a gradient in mass-to-light
ratio caused by induced star-formation during infall.  We note that
star-forming regions within galaxies clearly present no dynamical problems, but
also that large spirals would still appear as such in the HDF, which leads us
to favor a scenario in which the faint compact sources in the HDF are giant
starforming regions within small normal galaxies, such as Magellanic
irregulars.  Last we note that the ``excess'' number of correlated objects near
a given faint source approaches 1.5, suggesting that the previous counts of
objects have overestimated the number of galaxies by a factor of 2.5, while
underestimating their individual luminosities by the same factor.

\end{abstract}


\section{Introduction}

The Hubble Deep Field (Williams \etal\ 1995) affords us an unprecedented view
of the optical sky at small angular scales and faint flux levels.  The number
of sources detected in the field is a dramatic increase over previous faint
source counts and has led to proposals that the number of galaxies has been
seriously underestimated (Williams \etal\ 1996) by ground-based work.  However,
Colley \etal\ (1996b) (Paper I) have argued that many of the faintest objects
show strong angular correlation on the sub-arcsecond scale.  For this to be so,
many of the detected sources must lie within a few kiloparsecs of each other,
leading to the possibility that the excess is due to fragments of galaxies
mis-identified as separate objects.  We now seek to assess the dynamical nature
of these objects.

In section 2, we shall motivate the importance of understanding these
associated multiplets by computing the probability that an object's nearest
neighbor lies within one arcsecond ($\sim 6\kpc$), first using the correlation
function, then using nearest-neighbor statistics.

Our first dynamical hypothesis, which is favored in Paper I, is that these
physically associated objects are part of normal (in the present-day sense)
galaxies seen at high-redshift, under the peculiar observational effects of
strong $K$-correction and surface-brightness dimming.  These effects tend to
enhance strongly the relative prominence of UV-bright compact sources, such as
giant star-forming regions.  O'Connell \& Marcum (1996) have shown that even
nearby $L_*$ spirals observed in the ultraviolet appear very spotty and clumpy.
The light is dominated by active star-forming regions, while the underlying
stellar disk remains virtually invisible.  We showed in Paper I that the counts
are consistent with this hypothesis ($N_{objects} \propto 1/\mbox{flux}$), but
do not necessitate it.

We next consider the alternative hypothesis that these subgalaxian sources are
currently undergoing merging on their way to becoming present-day $L_*$
galaxies.  This hypothesis is probably the most exciting of those we will
consider, because it would suggest that we have finally seen deep enough to see
normal galaxies in formation.  This hypothesis has been proposed by several
others (\cf\ Odewahn \etal\ 1996), since large galaxies irregularly acquiring
gas on ``hot-spots'' where active star-formation ensues could explain the
multi-peaked appearance of sources in the HDF.

We finally discuss the possibility that the sources are $L_*$ galaxies
accreting small Irregulars on the dynamical friction timescale, much as our
galaxy is currently accreting the Magellanic Clouds.  Again, this hypothesis is
consistent with the results from Paper I, in that nearly complete accretion
events could show up as multiple peaks within a small angular (and spatial)
scale.

Ultimately we find that the first hypothesis is the most likely dynamically,
which indicates that the object counts have overestimated galaxy counts and
underestimated individual galaxy luminosities.

\section{Small Angle Correlation Among the Faint Sources}

In Paper I, we computed the angular correlation within several subsets of
objects observed in the HDF.  An important subset contained only objects with
high color-redshifts, i.e. objects having colors most consistent with
high-redshift ($z > 2.4$) populations (Steidel et al. 1996).  We found this
subset to contain excess correlation below $0.\arcsec 5$ as compared with the
complementary set of low color-redshift objects.  We plot in figure 1, for
reference, the correlation of these high color-redshift objects.  We have
overplotted the best-fit power law, which has the form $w_{fit} =
(\theta/\theta_\ell)^\alpha$, where $\alpha = -1.1 \pm 0.1$ and $\theta_\ell =
0.\arcsec 93 \times 10^{\pm 0.03}$, the correlation length.

We may now compute the expected number of objects within a correlation length
of a given object in the catalog as
\begin{equation}
\langle N(\theta<\theta_\ell) \rangle =
\int_0^{\theta_\ell}2\pi\Sigma(1+w)\theta d\theta =
\pi\Sigma \theta^2\left[1 +
{2\over{\alpha+2}}\left({\theta\over{\theta_\ell}}\right)^\alpha 
\right].
\end{equation}
where $\Sigma$ is the mean number surface density of catalog objects in the
field.  For our measured values of $\alpha$ and $\theta_\ell$, we have $\langle
N \rangle = 0.38$ (as compared to 0.12 with no correlation).  According to
Poisson statistics, $1-e^{-0.38} = 32\%$ of the catalog objects would have at
least one other object within the correlation length.

As a further check, we cataloged all objects in each chip which had a neighbor
within one arcsecond.  We found that 196 of the 695 objects classified as high
color-redshift (28\%) had at least one neighbor within one arcsecond.
Including multiplets, we have 256 total objects within an arcsecond of another
($\langle N \rangle = 0.37$), in excellent agreement with the values expected
from the correlation.  Also note that the directly counted number of associated
objects is roughly a factor of three greater than that expected from a random
distribution.

We have plotted in figure 2a the distribution of the nearest neighbors.  The
solid histogram is the differential distribution of nearest neighbors
(separation $\theta_{nn}$) among high color-redshift catalog objects in the
HDF.  We have plotted for reference the expected distribution for a random
sample with the same number of objects in the same field (dashed curve)
\begin{equation}
{{dP}\over{d\theta_{nn}}} = e^{-N(\theta_{nn})}{{dN}\over{d\theta_{nn}}},
~~~~ dN(\theta_{nn}) = 2\pi\Sigma\theta_{nn}d\theta_{nn}.
\end{equation}
where $\Sigma$, as before, is the number surface density of catalog objects.
The peak in the distribution of catalog objects occurs between $0.\arcsec 5$
and $1.\arcsec 0$, while in a random sample, the distribution of nearest
neighbors would be expected to peak near $1\arcsec .6$.  (As an aside we note
that this result could not have been predicted simply from the correlation
function.  All order, not just second order, correlations are needed to predict
the nearest neighbor distribution.)

Figure 2b shows the same data in a different way.  In this figure we plot the
number of objects within an angular radius $\Delta\theta$ of other objects.
The solid curve is for the high color-redshift objects in the HDF, the dashed
curve for a random distribution.  The heavy solid curve is the excess over the
random distribution.  We see that the significant correlation function actually
translates into the number of objects expected from the correlation, denoted by
the plus-sign, and that the excess over the random sample is significant with
an arcsecond, and approaches 1.5 objects at large separation.

\section{Hypothesis 1: The sources are parts of Normal Galaxies}

We shall first consider this basic question: What would normal present-day
galaxies look like if redshifted to $z > 1$?  As discussed in Paper I, two
observational effects critically change the appearance of galaxies as they are
redshifted away from us.  The $K$-correction brings rest-frame ultraviolet
light into the visible bands of the HDF, and surface brightness dimming favors
prominence of compact (unresolved or marginally resolved) objects.  Giant \Hii\
regions are both UV bright and compact, and hence shine brightly above the
underlying diffuse stellar disk.  The $K$-correction effects have been well
established by O'Connell \& Marcum (1996) who have imaged nearby spirals with
UIT and shown them to present a very clumpy appearance where even the bulge can
be less prominent than the active star-forming regions in the disk.  This
effect is exacerbated by preferential surface brightness dimming of diffuse
objects with increasing redshift.

An excellent nearby example of a bright star-forming region in a diffuse
stellar disk is 30 Doradus in the Large Magellanic Cloud (LMC).  Cheng \etal\
(1992) have measured the UV ($2558\AA$) flux of the inner $3^\prime$ (40 pc) at
6.0 magnitudes ($M_{UV} = -12.7$).  The total 30 Doradus complex, however
covers $1^\circ$ (1 kpc), so that the total absolute $U$-magnitude of the
entire complex is approximately $-14$.  For a distance modulus of 43.7 ($z \sim
1$, $h = 0.7$, Peebles 1993), this gives a magnitude of approximately 29.7,
within the broad peak of the $(R+I)/2$ magnitudes of objects in the Hubble Deep
Field (Paper I).  ``Super star clusters'' (as seen in starburst galaxies) can
have somewhat brighter UV flux (up to around 3 magnitudes brighter, O'Connell
\etal 1995).  Therefore, a large spread in the magnitude around 28--29 for very
actively star-forming regions in the Hubble Deep Field would be expected, and
is confirmed in Paper I.  At a redshift of $z = 1.0$ the 0.5 kpc radius of 30
Doradus translates to an angular radius of 0.08 arcseconds, or $80\%$ of one
p.s.f. on the WFC2.  Thus we might expect many of the small objects seen in the
HDF to be marginally resolved if our hypothesis were correct.

The relative $K$-correction between such star-forming regions and diffuse
stellar disks may be quantified with stellar population synthesis models.
Charlot \& Bruzual (1995) report a $(\lambda = 2700\AA) - V$ color of order
$-1.5$ in regions which have undergone a burst of star-formation $10^7$ years
ago, vs. $-0.1$ for a $3~\mbox{Gyr-old}$ population in an exponentially
declining star-formation epoch. For an old stellar disk in the same model (age
of 4~$\mbox{Gyr}$), the color is 1.0.  So at a redshift $z \sim 1.5$, a young
\Hii\ region will be enhanced over a Population I disk by about 1.5 magnitudes
due to $K$-correction alone, while the enhancement over a Population II bulge
can be as great as 2.5 magnitudes.

Furthermore, while a large fraction of the light from a large star-forming
region resides in one $0.1\arcsec$ point-spread-function of the WFC2, the
diffuse light of given surface magnitude only lends one-hundredth of that light
to a single pixel.  So, in comparison to ground-based efforts, objects which
are marginally resolved from space can see a several magnitude enhancement in
surface brightness over diffuse objects.  Finally, fully resolved objects (such
as diffuse disks) will suffer $(1+z)^4$ dimming in bolometric surface
brightness, significant particularly for higher redshift objects.

All these factors transform the surface brightness from the normal level of 21
mag/sq. arcsec to 31 mag/pixel, significantly dimmer than the star-forming
regions, and close to the detection threshold of the HDF.  Also, nearly all
local, late-type galaxies achieve a maximum a B-surface brightness of 21 or
greater in their disks (McGaugh, 1996; Patterson \& Thuan 1996).  Among a
significant fraction of the associated faint sources, very faint (often just
detectable) material connecting them is visible (\cf\ Steidel \etal\ 1996), in
agreement with the hypothesis that the objects are physically associated within
some underlying background medium.

This rough quantitative sketch demonstrates that giant star-forming regions
posess the necessary properties to constitute many of the faint sources in the
HDF, while the underlying disks would be sufficiently dimmed from
$K$-correction and surface brightness dimming to push them to the edge of
detectability on the HDF.

In fact, we may be more specific about the nature of the hosts for these
star-forming regions.  O'Connell and Marcum (1996) have produced synthetic
images of $L_*$ spirals at high-redshift with the approximate resolution of the
HDF.  One could summarize the results by saying that, although they look a lot
more clumpy, spirals still look like spirals.  In M101, for instance, the \Hii\
regions can be traced around the spiral arms quite easily in these synthetic
images.  Since 1) we masked out obvious large spirals in our study (see Paper
I), 2) most of the objects in our high color-redshift catalog have only two or
three peaks (consistent with the analysis of the correlation in the previous
section), and 3) the bulk of the excess correlation occurs within $6\kpc$, we
see that the observational evidence suggests that the objects are most akin to
Magellanic irregulars, which contain one or more very bright Giant \Hii\
regions, such as 30 Doradus, that dominate the flux seen in the HDF.  Obviously
this is dynamically plausible, since the LMC exists.

\section{Hypothesis 2: The sources are Large Galaxies in Formation}

In order to model the dynamics for other theoretical hypotheses, we require a
mass-model for the typical multiple source system.  Assuming a constant
mass-to-light ratio (reasonable for coeval active star forming regions), we
directly infer the mass surface density from the surface brightness, which we
have plotted in figure 3.  Figure 3 plots the $(R+I)/2$ surface magnitude in
catalogued sources within bins of logarithmic angular radius about other
sources.  This process averages the profile of light in all sources with
respect to each other and is independent of choosing a ``central object,''
which would be a daunting task for most of these rather irregular objects.  We
see that the profile divides roughly into two power-laws, a steep power-law
(slope $= -2.9 \pm 0.6$) inside of one arcsecond, and a much shallower one
outside (slope $= -0.5 \pm 0.2$).  This plot alone is suggestive that the
dynamics inside of one arcsecond may be different from those outside, which one
would expect if the multiple sources are dynamically associated within $6\kpc$,
as suggested in Paper I.  The slopes from figure 3 will be critical in our
consideration of the dynamical consequences of our hypotheses.

The most extreme dynamical hypothesis is that the HDF objects are
self-gravitating objects merging into larger systems. It has been proposed
recently (Burkey et al. 1994) that many faint pairs of objects seen by \HST\
are undergoing mergers.  The current status of the extreme merger scenario is
outlined in Carlberg (1992).

To explore this hypothesis, we consider typical faint HDF sources with visual
magnitude of 28.8, or $M = -14.9$ with the adopted distance modulus for
redshift $z \ga 1$.  The $K$-correction will shift the observed light into the
$B$ and $U$ passbands where $M_\odot \approx 5.5$.  Thus the UV luminosity of
the median object is $L_{med} = 1.4\times 10^8L_\odot$.  If these objects are
really infalling dwarf galaxies, their mass-to-light ratio, ($\Upsilon$),
should be characteristic of a stellar population 3--4~Gyr old.  The population
synthesis models of Worthey (1994) yield $\Upsilon_B = 2.2$ for $t =
3~\mbox{Gyr}$, bringing the mass of the objects to $3.2\times 10^8M_\odot$.  We
will adopt this conservative estimate of $\Upsilon$---should dark matter
contribute significantly to the mass, the corresponding increase in $\Upsilon$
would strengthen our conclusions.

The surface brightness profile of the multiple-peak HDF objects inside a
projected radius of $10\kpc$ can be approximated, as discussed above, by a
power law $\Sigma(R) = \Sigma_0(R/R_1)^{-\alpha}$, with $\alpha \approx 2.9$,
$\Sigma_0 = 6\times 10^7L_\odot\cdot\mbox{arcsec}^{-2}$ and $R_1 = 1\arcsec$
(5.9 kpc with $z\sim 1$ and $h = 0.7$).  Inside $0.\arcsec 4$ there are no
pairs of objects, which is likely an artifact of our object detection
algorithm, which requires some smoothing, and distinct separation (see Paper
I).  Since the slope of the light in individual (but associated) sources is
fairly constant down to this radius, we will consider $r_{in} = 0\arcsec.4$
($\sim 2.3\kpc$) as the radius within which accreted material must have settled
(we will see that a smaller radius would only require more central density, so
this estimate is also conservative).

Solving the Abel equation (Binney \& Tremaine 1987) we obtain the spherically
symmetric distribution of objects in three dimensions
\begin{equation}
\rho(r) = {\alpha I(\alpha) \Upsilon \Sigma_0 \over \pi R_1} \;
   \left( {r \over R_1} \right)^{-\alpha-1},
\end{equation}
where $I(\alpha)$ is a dimensionless integral that depends weakly on
the parameter $\alpha$. $I(\alpha=2.9) \approx 0.7$. The total mass
inside the fiducial radius $R_1$ integrates to
$M(R_1) = 1.6\times 10^9\, M_{\sun}$,
Although each particular system that we are observing may have a short life
time, it is reasonable to assume that the observed distribution of separations
persists for a long enough time, say the Hubble time at that redshift, to be in
a quasi-steady state.  This enables us to estimate the mass of the central
system. From the continuity equation, the mass accretion rate in a spherical
system at a radius $r_{in}$ is
\begin{equation}
\dot{M}_{acc} = 4\pi r_{in}^2 \rho{(r_{in})} v_r,
\label{eq:cont}
\end{equation}
where $\rho{(r_{in})}$ is the deprojected space density of the objects, and
$v_r$ is the radial infall velocity. We assume for simplicity that all objects
are falling in on radial orbits with the speed determined by the central mass,
i.e. $v_r = (GM_{acc}/r_{in})^{1/2}$.  While this assumption is somewhat
extreme, the order-of-magnitude of the velocity will be consistent with that
used in the general merger scenario.  Choosing $r_{in} = 0.\arcsec 4 \approx
2.3\kpc$, we can integrate equation (4) to obtain the total accreted mass
within $r_{in}$ 
\begin{equation}
M_{acc}(t) =
5\times10^{12}\left({{t}\over{3\times 10^9\mbox{yr}}}\right)^2M_\odot,
\end{equation}
by an age of 3 billion years, the Hubble time for $z \sim 1\mbox{--}2$. This
much mass inside the inner 2 kpc is obviously in contradiction with $z = 0$
observations. Therefore this extreme merger scenario is not supported by the
HDF data.

Another test for the merger hypothesis is the total merger rate over the Hubble
time and the corresponding mass density of the merger remnants.  We have found
695 objects in three WFC2 chips of $72\arcsec \times 72\arcsec$.  Since the
total radial (redshift) extent of these objects is uncertain, we take it simply
as the Hubble distance, $3\times 10^3 \mbox{Mpc}$, times some factor $\chi <
1$.  By redshift $z\sim 1$, we expect to see several merger remnants in each
cubic megaparsec.

\begin{equation}
n_{merger} =
2.5\left({h\over{0.7}}\right)
\left({\chi\over{1/3}}\right)^{-1}
\mbox{Mpc}^{-3}.
\end{equation}

If each merger accumulates $5\times 10^{12}M_\odot$, the total amount of mass
density grossly exceeds the critical density of the Universe, and is excluded
by local ($z = 0$) dynamical measurements;

\begin{equation}
\Omega_{merger}\equiv{{\rho_{merger}}\over{\rho_{cr}}} =
11\left({h\over{0.7}}\right)
\left({\chi\over{1/3}}\right)^{-1}~~\mbox{for}~~\Lambda = 0.
\end{equation}

\section{Hypothesis 3: The sources are Accreting Satellites of a Normal Galaxy}

Another initially plausible hypothesis is that the faint objects might be
satellites accreting onto normal $L_*$ galaxies due to dynamical friction
within the dark matter halo. This situation is a clear analogue of the
Magellanic Clouds and the Sagittarius dwarf galaxy around the Milky Way.

Assuming the isothermal distribution of the dark halo with
the rotation speed of 220 km s$^{-1}$, the infall time from $r$ to the
center due to dynamical friction is (Binney \& Tremaine, 1987)
\begin{equation}
t_{df} = 1.0\times 10^9\; \left( {r \over 5.9\ {\rm kpc}} \right)^2 \;
  \left( {v_c \over 220\ {\rm km\ s}^{-1}} \right)\; \left( {3.2\times 10^8\,
  M_{\sun} \over M} \right)\; {\rm yr}.
\end{equation}
This timescale is still short enough so that all presently observed objects
have enough time to sink into the center.

Dynamical friction can either deplete the initial population of objects or
increase it, depending on the initial radial distribution of accreted objects.
Ostriker \& Turner (1979) showed that if the space number density of the
objects is inversely proportional to the first power of the distance from the
center, $r$, a steady state is achieved, because at any radius, $r$, as many
objects sink inward as come from outside.  One can write the continuity
equation for the space density, $\rho$, in objects as
\begin{equation}
{{\partial\rho}\over{\partial t}} = {1\over{r^2}}
{{\partial}\over{\partial r}}\left(r^2\rho v_r\right).
\end{equation}
Substituting for $v_r = -r/t_{df} \propto r^{-1}$, and $\rho \propto
r^{-\alpha}$, we have
\begin{equation}
 {{\partial\rho}\over{\partial t}} \propto (\alpha - 1)\times r^{-\alpha-2}.
\end{equation}
The steady-state solution at $\alpha = 1$ is apparent.  If the original value
of $\alpha$ is greater than one, then a core with slope $\alpha = 1$ develops
within about the dynamical friction time at a given radius.  A shelf develops
at the interface between the $\alpha = 1$ core and the $\alpha > 1$ outer
regions.  If the $\alpha$ is less than one, the interior also develops a slope
of about $\alpha = 1$ within a dynamical friction time, but the interface shows
a dip between the outer shallow slope region and the inner region of greater
slope.  Either way, the system will eventually relax into an $\alpha = 1$
steady-state throughout until all the mass has been accreted.

In figure 3, we have plotted the best-fit power law to the surface density
distribution, which corresponds to a spatial density decreasing with a
power-law slope of $\alpha = 3.9 \pm 0.6$ inside one arcsecond.  For $\alpha =
1$, we would expect the surface density to be decreasing only logarithmically
with radius.  While the data are not perfect, they obviously do not portray
this very weak variation expected from dynamical friction evolution.
Furthermore, inside one arcsecond, the $\chi^2$ value for the best-fit slope is
0.98 per degree of freedom, but 15 per d.o.f. for the $\alpha = 1$ case.

We now take the slope of the observed space distribution of the faint sources
in the HDF within one arcsecond ($5.9\kpc$) to be $\approx -3.9$.  Since the
dynamical friction time at this radius is roughly one billion years, we expect
to be observing such objects already in a steady-state with $\alpha = 1$ by $z
\la 5$ if accretion is important.  Thus the slope of $\alpha = 3.9$ inside one
arcsecond is a contradiction to the expected slope.

\section{Possible Effects of a Luminosity Gradient}

In hypotheses 2 and 3 above, we have assumed that the excess in light and in
number counts inside $1\arcsec$ corresponds directly to an excess in mass.
However, one could imagine that tidal shocks and other dynamical stimuli in
infalling matter might induce star-formation there, and thus reduce the
mass-to-light ratio significantly with decreasing separation.

First, we consider the plausible range of $\Upsilon$ as given by 30 Doradus at
the low end, and the LMC as a whole at the high end.  30 Doradus has a
mass-to-light ratio of $\Upsilon_{min} \approx 0.01$, while the LMC as a whole
has $\Upsilon_{max} \approx 20$.  In our merger calculation, we assumed
$\Upsilon = 2.2$, which is conservative, since LMC is, itself, undergoing shock
induced star-formation at the hands of our galaxy's tidal field.  Moreover, it
would be hard to imagine that if these faint objects are really infalling
merger fragments that they would be devoid of dark matter completely.

We address the possibility of a gradient in the mass-to-light ratio by
considering relative slopes in number density and luminosity density as a
function of radius.  First, outside of $1\arcsec$, both the number and
luminosity density go as $r^{-1.4}$, which implies a constant mass-to-light
ratio, if the infalling objects have the same mass.  If we assumed that there
were no real mass excess, the number density would maintain the same power-law
all the way in.  To explain the observed excess in light (figure 3), one would
require the mass-to-light ratio to decrease as $\Upsilon \propto r^{2.6}$ with
decreasing radius.  This implies a change in mass-to-light of order 20 from
$2.4\kpc$ to $8\kpc$.  This is a large change, but not impossible if
significant star-formation is induced by tidal shocks.

However, under this assumption, we must check whether a varying mass-to-light
is sufficient to explain the luminosity excess with no mass excess.  To do so,
we imagine brightening the luminosity function with decreasing radius to pull
less massive objects over the flux limit.  We take the number-magnitude
relation observed in Paper I as $n(>L_{limit}) \propto L_{limit}^{0.25}$ (which
has a convergent integrated flux), and recall from figure 3 that $L \propto
r^{-3.9}$, so that $n(>L_{limit}) \propto r^{-1}$.  However, this disagrees
with the observed number density, which has slope $n \propto r^{-3}$.  The
number density is significantly steeper than we would expect under this
hypothesis, that induced star-formation in approaching objects explains the
steep inner slope of the luminosity density.  We therefore conclude that there
is a real mass excess which explains the bulk of the luminosity correlation in
figure 3.

Though physically motivated, and plausible, a variable $M/L$ caused by
shock-induced star-formation cannot account for the number and light excesses
observed in figures 1--3, and thus our conclusions are not significantly
affected by this possibility.

\section{Conclusions}

In Paper I, we discovered that the faint, high color-redshift objects detected
in the Hubble Deep Field present an angular correlation which is significant
at angular scales below one arcsecond ($\sim 6~\mbox{kpc}$).  In this work, we
have used the correlation function and nearest neighbor statistics to determine
that roughly one-third of all the selected objects have a neighbor within one
arcsecond, so that dynamical interactions between them must be important.

We then examined three hypotheses as to the dynamics of the faint multi-peaked
sources in the Hubble Deep Field.  We find that the most likely scenario is
that the sources are giant star-forming regions which reside in Magellanic
Irregulars in approximate steady-state.

This hypothesis is favored on several grounds.  From an observational
standpoint, we find that the luminosity of Giant \Hii\ regions, such as 30
Doradus and those in other local galaxies, have UV luminosities roughly
consistent with the bulk of faint sources observed in the HDF, yet few if any
high-redshift $L_*$ spirals are visible in the HDF.  Also, the disk of LMC
would be just detectable if observed at high-redshift in the HDF, which is
consistent with the faint emission seen around some of the multi-peaked sources
in the HDF.  Finally, the size of giant \Hii\ regions corresponds to the barely
resolved apparent size of the ``small galaxies.''  On more theoretical grounds,
dynamical arguments make quite unlikely a scenario in which the objects are
infalling fragments merging into an $L_*$ galaxy, as such a system would
produce too much mass ($\sim 5\times 10^{12} M_\odot$) inside $2\kpc$ by $z
\sim 1$.  Furthermore, $L_*$ spirals with orbiting satellites could not be in a
steady-state with their observed number density slope.  The steady-state slope
of $\rho \propto r^{-1}$ should be achieved at in roughly a dynamical time,
which is $1\Gyr$ at one arcsecond; instead we see a slope of $\rho \propto
r^{-3.9}$.  Finally, we find that it is not possible to simultaneously explain
the small angle number excess and luminosity excess with a single variable
mass-to-light ratio.

We therefore suggest that very bright active starforming regions within
Magellanic dwarf irregulars at $z \ga 1$ provide the bulk of the physically
associated faint blue sources in the Hubble Deep Field.  As a consequence,
number counts can overestimate the number of galaxies by a factor of 2.5 and
underestimate the individual luminosities of galaxies by the same factor.

The most straight-forward test of our favored hypothesis is to study the field
with longer exposures at longer wavelengths where the underlying stellar disks
should be most visible.  Also, we predict a relatively small relative velocity
difference between adjacent peaks at the level of $\la 50~\mbox{km/s}$, far
below the level of $\ga 200~\mbox{km/s}$ expected for objects which are
actively merging.  One recent study (Guzm\'an \etal\ 1997, submitted during the
refereeing stage of this paper), has, in fact, verified that many of the faint
objects in the HDF are undergoing rapid star-formation, as inferred from
significant emission in lines.  Moreover, recent results from gravitational
lens reconstruction of high-redshift galaxies has revealed sources which
have multiple intense star-forming regions within a several kiloparsec disc
(Franx \etal\ 1997, Colley \etal\ 1996a).

\begin{acknowledgements}

We thank an anonymous referee for useful ideas in strengthening our
conclusions.  WNC is most grateful for the continued support of the Fannie and
John Hertz Foundation, and partial support from NSF grant AST-9529120. OYG and
JPO's work has been partially supported by NSF grant AST-9424416, and by NASA
grant NAG5-2759.  JER's work has been supported by NSF grants AST 91-17388,
NASA ADP grant 5-2567, and JER's NSF traineeship DGE-9354937.  Finally, we very
kindly thank the HDF team for their hard work and generosity in preparing the
data for public release.

\end{acknowledgements}

\begin{figure}[p]
\plotfiddle{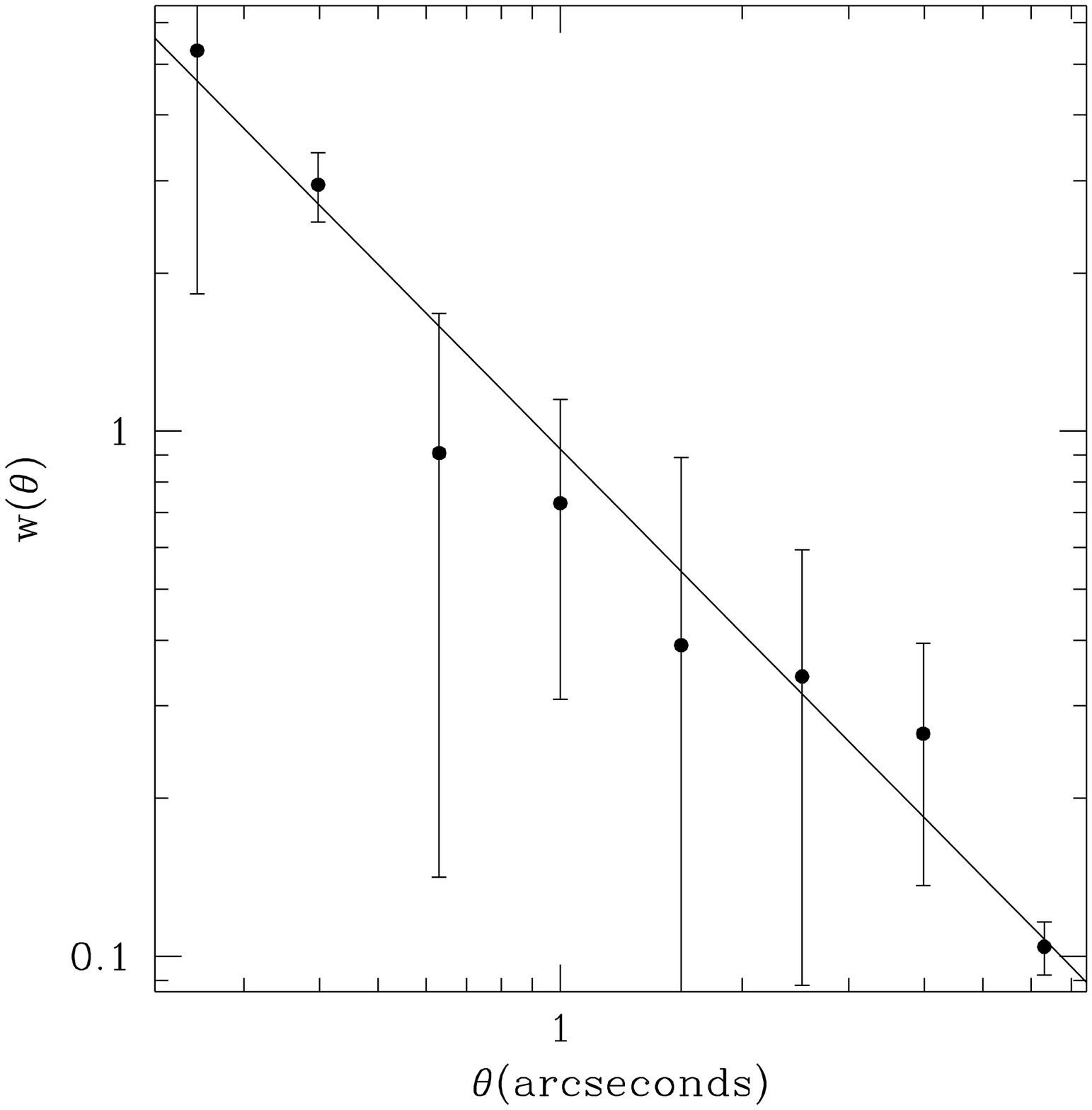}{12cm}{0}{60}{60}{-185}{-90}
\caption{ The angular correlation function of high color-redshift objects in
the Hubble Deep Field.  Overplotted is the best-fit power-law to the
correlation, $w_{fit} = (\theta/\theta_\ell)^\alpha$, where $\alpha = -1.1
\pm 0.1$ and $\theta_\ell = 0.\arcsec 89 \times 10^{\pm 0.03}$}
\end{figure}

\begin{figure}[p]
\plotfiddle{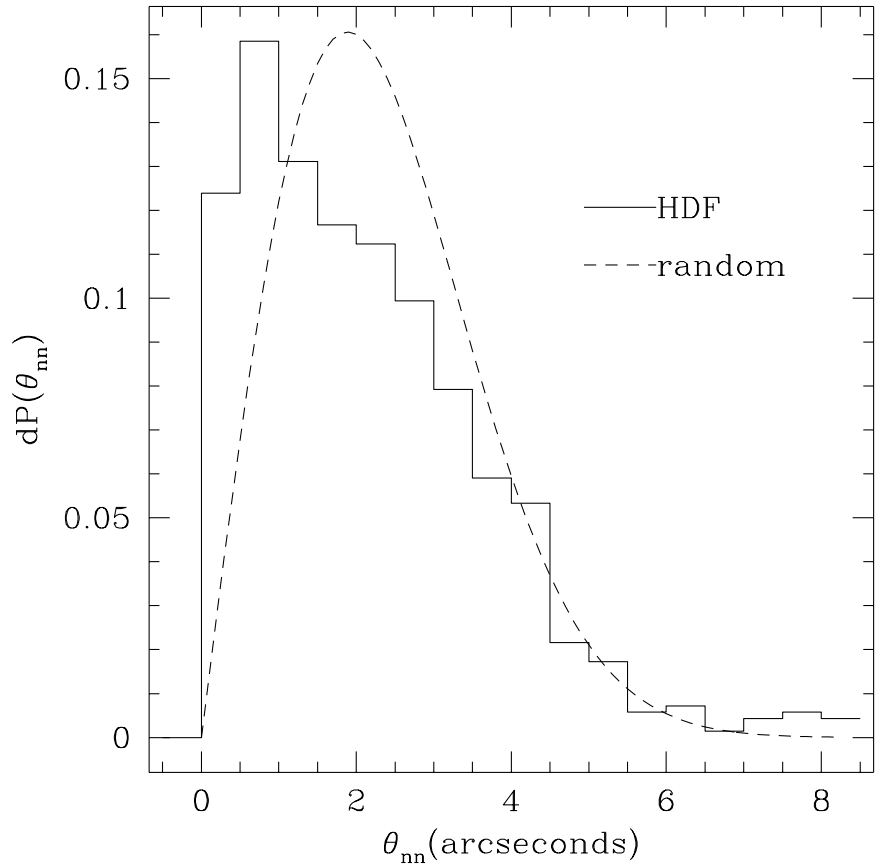}{12cm}{0}{60}{60}{-185}{-90}
\caption{a) Distribution of nearest neighbors among high color-redshift catalog
objects in the Hubble Deep Field (solid) and for a random distribution with the
same number of particles (dashed).  Where $\theta_{nn}$ is the angular
separation of an object and its nearest neighbor, we have plotted both the
differential distribution, $P(\theta_{nn})$.}
\end{figure}

\begin{figure}[p]
\addtocounter{figure}{-1}
\plotfiddle{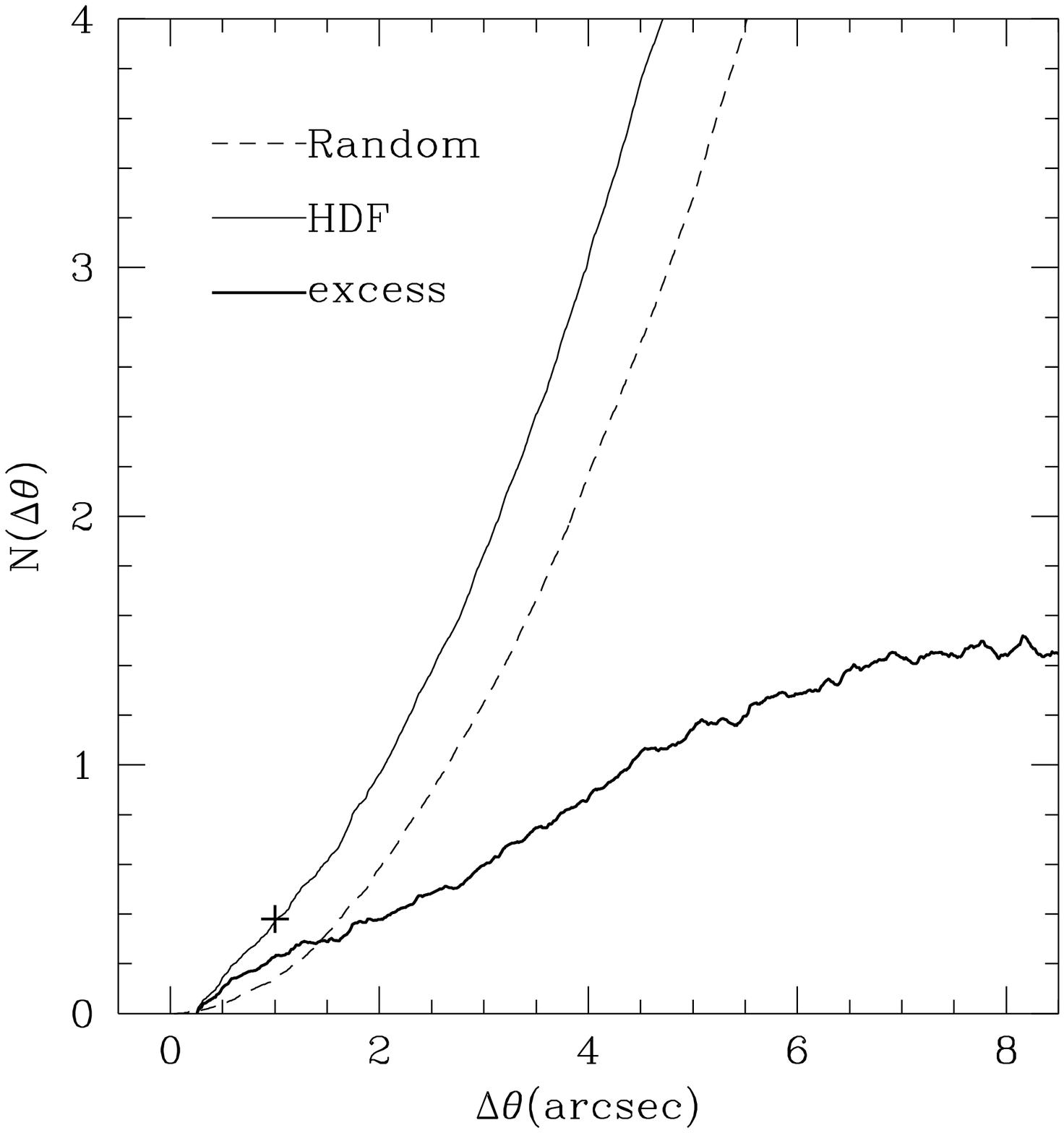}{12cm}{0}{60}{60}{-185}{-90}
\caption{b) Cumulative distribution of all neighbors of high color-redshift
objects in the Hubble Deep Field (light solid curve), the same for a random
distribution (dashed curve), and the excess in the HDF (heavy solid line).  The
large plus sign denotes the expected number derived from the correlation
function.}
\end{figure}

\begin{figure}[p]
\plotfiddle{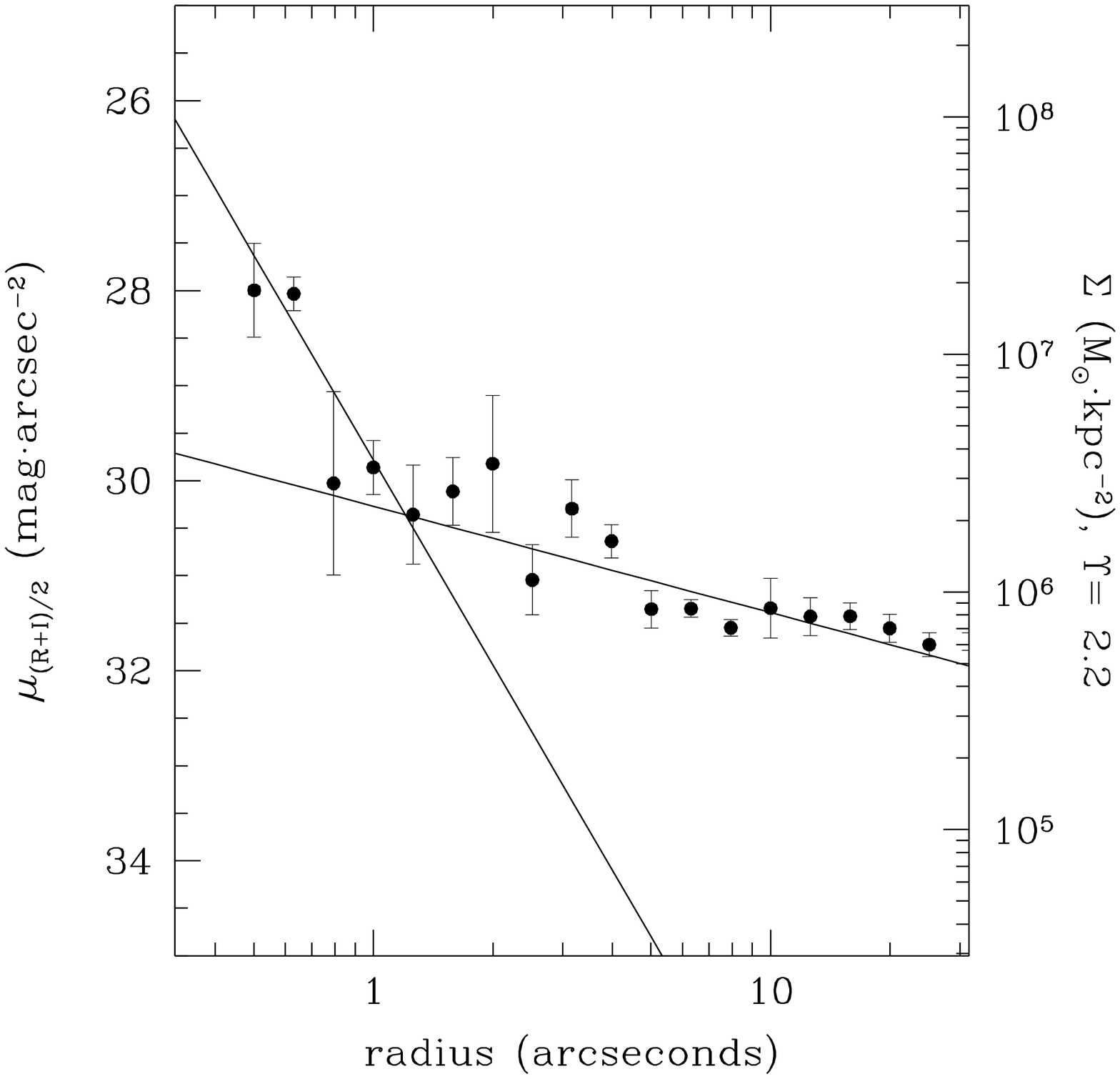}{12cm}{0}{60}{60}{-185}{-90}
\caption{ The magnitude and mass (assuming mass-to-light, $\Upsilon = 2.2$ and
distance modulus = 43.9 for $z \sim 1$) in catalog objects as a function of
logarithmic radius.  Surface magnitude is that measured in individual catalog
sources.  Overplotted are two power-law fits (slope = $-2.9\pm 0.6$ inside one
arcsecond; slope = $-0.5 \pm 0.2$ outside one arcsecond), from which the
three-dimensional mass density in the luminous sources is derived (see Text).}
\end{figure}


\begin{references}

\reference{}Binney, J.J., \& Tremaine, S. 1987, {\it Galactic Dynamics\/}  
(Princeton University Press: Princeton)

\reference{}Burkey, J. M., Keel, W. C., Windhorst, R. A., \& Franklin, B. E.
1994, \apjl, 429, L13

\reference{}Charlot, S., \& Bruzual, G.A., 1995, results from ``galaxevpl.f''
FORTRAN program

\reference{}Carlberg, R. G. 1992, \apjl, 399, L31

\reference{}Cheng, K., \etal, 1992, \apj, 395, L29

\reference{}Colley, W., Rhoads, J.E., \& Ostriker, J.P., \& Spergel,
D.N., 1996a, \apj, 473, L63 (Paper I) 

\reference{}Colley, W., Tyson, J.A., \& Turner, E.L., 1996b, \apj, 461, L83

\reference{}Franx, M., Illingworth, G.D., Kelson, D.D., van Dokkum, P.G., Tran,
K., 1997, astro-ph/9704090

\reference{}Kron, R. G., 1995, in The Deep Universe, ed. Sandage, A. R., Kron,
R. G. and Longair, M. S., (Berlin: Springer-Verlag)

\reference{}McGaugh, 1996, \mnras, 280, 337

\reference{}Nicolet, B., 1978, A\&AS, 34, 1

\reference{}O'Connell, R. W. \& Marcum, P., 1996, ``The Ultraviolet Morphology
of Galaxies'' in {\it HST and the High Redshift Universe (37th Herstmonceux
Conference)}, eds. N.R. Tanvir, A. Aragon-Salamanca, J.V. Wall, 1996.

\reference{}O'Connell, R. W., Gallagher, J. S., Hunter, D. A. \& Colley, W. N.,
1995, \apjl, 446: L1

\reference{}Odewahn, S. C., Windhorst, R. A., Driver, Simon P. \& Keel, W. C.,
1996, \apj, 472, L13

\reference{}Ostriker, J.P., \& Turner, E.L., 1979, \apj, 234, 785

\reference{}Patterson, R.J., \& Thuan, T.X., 1996, \apjs, 107, 103

\reference{}Peebles, P.J.E., 1993, Principles of Physical Cosmology, Princeton
University, Princeton

\reference{}Guzm\'an, R., Gallego, J., Koo, D. C., Phillips, A. C., Lowenthal
J. D., Faber,  S. M., Illingworth, G. D., \& Vogt, N. P., astro-ph/9704001

\reference{}Space Telescope Science Institute, 1995, ``Filter Selection
for the Hubble Deep Field'' (Baltimore: STScI)

% \reference{}Steidel, C., 1996a, AJ, in press
\reference{}Steidel, C., Giavalisco, M., Dickinson, M.,
 \& Adelberger, K. 1996, AJ, in press

\reference{}Williams, R.E., Blacker, B.S., Dickinson, M., Ferguson, H.C.,
Fruchter, A.S., Giavalisco, M., Gilliland, R.L., Lucas, R.A., McElroy, D.B.,
Petro, L.D., \& Postman, M., ``The Hubble Deep Field Observations,'' 1995, in
Science with the Hubble Space Telescope---II, P. Benvenuti, F. D. Macchetto, \&
E. J. Schreier, eds.  (Baltimore: STScI)

\reference{}Williams, R.E. \etal\ 1996, \aj, 112, 1335

\reference{}Worthey, G. 1994, \apjs, 95, 107

\end{references}
\end{document}